# *Ab initio* melting temperatures of bcc and hcp iron under the Earth's inner core condition

Yang Sun[1,2,3*], Mikhail I. Mendelev[4], Feng Zhang[3], Xun Liu[5], Bo Da[5], Cai-Zhuang Wang[3]

Renata M. Wentzcovitch[2,6,7] and Kai-Ming Ho[3]

[1]Department of Physics, Xiamen University, Xiamen, Fujian 361005, China
[2]Department of Applied Physics and Applied Mathematics, Columbia University, New York, NY 10027, USA
[3]Department of Physics, Iowa State University, Ames, IA 50011, USA
[4]Intelligent Systems Division, NASA Ames Research Center, Moffett Field, CA 94035, USA
[5]Research and Services Division of Materials Data and Integrated System, National Institute for Materials Science, Ibaraki 305-0044, Japan.
[6]Department of Earth and Environmental Sciences, Columbia University, New York, NY 10027, USA
[7]Lamont–Doherty Earth Observatory, Columbia University, Palisades, NY 10964, USA

**Abstract**

There has been a long debate on the stable phase of iron under the Earth's inner core conditions. Because of the solid-liquid coexistence at the inner core boundary, the thermodynamic stability of solid phases directly relates to their melting temperatures, which remains considerable uncertainty. In the present study, we utilized a semi-empirical potential fitted to high-temperature *ab initio* data to perform a thermodynamic integration from classical systems described by this potential to *ab initio* systems. This method provides a smooth path for thermodynamic integration and significantly reduces the uncertainty caused by the finite-size effect. Our results suggest the hcp phase is the stable phase of pure iron under the inner core conditions, while the free energy difference between the hcp and bcc phases is tiny, on the order of 10s meV/atom near the melting temperature.

# 1 Introduction

Earth contains a solid inner core and a liquid outer core, both dominated by iron. The solid iron is generally believed to have the hexagonal close-packed (hcp) structure under the inner core conditions, while the body-centered cubic (bcc) structure is also suggested to be stable, especially when alloying with light elements (Belonoshko et al., 2017; Dubrovinsky et al., 2007; Kádas et al., 2009; Matsui and Anderson, 1997; Ross et al., 1990; Vočadlo et al., 2003). The competition between the hcp and bcc polymorphs mainly depends on their Gibbs free energy near the melting temperatures ($T_m$) under inner-core pressures. Because of the solid-liquid coexistence at the inner-core boundary, knowledge of the $T_m$ of competing crystal phases is crucial for understanding the thermodynamical stability of the inner core's structure (Alfè, 2009; Alfè et al., 2002b; Belonoshko et al., 2000; Laio et al., 2000; Sola and Alfè, 2009; Sun et al., 2018, 2022; Zhang and Lin, 2022). A solid phase with the highest melting temperatures is the stable phase, while other solid phases are metastable around $T_m$. However, the hcp and bcc iron's melting temperatures under Earth's core pressures are not well constrained. Experimental measurement of this quantity is challenging due to the difficulties in generating extreme conditions and detecting melts, which has caused different $T_m$ results ranging from 4,850 K to 7,600 K (Fischer, 2016). Most recent experiments have better constraints on melting temperatures but still have an uncertainty of ~500 K, which can be even more significant than the $T_m$ difference between the bcc and hcp phases (Anzellini et al., 2013; Hou et al., 2021; Kraus et al., 2022; Li et al., 2020; Sinmyo et al., 2019; Turneaure et al., 2020). Besides, differentiation between bcc and hcp structure in high P-T experiments is non-trivial, so bcc iron has rarely been reported at core pressures and temperature (Hrubiak et al., 2018).

Computer simulation provides an alternative way to access extreme P-T conditions and measure the properties of iron in the Earth's interiors. Depending on the description of atomic interaction, these simulations can be classified as classical or *ab initio* simulations. The classical molecular dynamics (CMD) simulations utilize semi-empirical potentials of the interatomic interaction. They can simulate large length-scale atomic structures for a relatively long time (e.g., million atoms for microseconds). This ability allows the implementation of the solid-liquid coexistence (SLC) approach to measure the melting temperature (Morris et al., 1994). The melting temperature calculated with this approach does not involve any approximations (other than the potential itself) and can be easily determined with uncertainty not larger than $0.0005 T_m$ (Wilson et al., 2015). However, the limitation of this method is that the $T_m$ highly depends on the accuracy of the employed semi-empirical potential. CMD simulations with different EAM potentials have reported significantly different $T_m$ for the bcc phase at inner core pressures, e.g., 7197 K in (Belonoshko et al., 2021) while 5974 K in (Sun et al., 2022) at 360 GPa. On the contrary, *ab initio* simulations provide accurate descriptions of the atomic interaction based on the electronic structure calculations with density functional theory (DFT). However, due to the time and length scale limitations of DFT, SLC simulations with *ab initio* calculations are usually computationally expensive and involve large uncertainty. For instance, previous SLC simulations with 256 Fe atoms lead to an uncertainty of ~ 500 K, which makes bcc and hcp's $T_m$ indistinguishable (Bouchet et al., 2013). The lowest uncertainty of iron's $T_m$ from *ab initio* SLC was ~100 K, achieved by Alfè using 980 atoms for hcp phases (Alfè, 2009). Unfortunately, the bcc phase was not studied.

Compared to SLC, the free energy approach is more widely used to measure the melting temperature from *ab initio* simulations. It is based on explicit calculations of the Gibbs free energy of solid and liquid phases, usually involving thermodynamic integration (TI). TI provides the free energy difference between the target and reference systems for which the absolute free energy is known *a priori*. The harmonic crystal and the ideal gas or simple liquids are popular choices as the reference states for solids and liquids, respectively (Frenkel and Ladd, 1984; Menon et al., 2021). While TI provides a decent estimate for absolute free energies, a relatively small inaccuracy in the solid-liquid free energy difference can significantly affect $T_m$ (Gunawardana et al., 2014). It was estimated that an uncertainty of 10 meV/atom in the free energy difference could lead to an uncertainty in $T_m$ of ~100 K (Alfè et al., 2001). So far, the free energy calculations and corresponding $T_m$ were available for hcp Fe, with uncertainties of 100-200 K (Alfè et al., 2002b; Sun et al., 2018). The Helmholtz free energy of bcc phase was demonstrated to be higher than hcp (Vočadlo et al., 2003), while bcc's $T_m$ has not been computed by TI yet.

In principle, one does not need the absolute free energy value to obtain the melting temperature. It is the *free energy difference* between the liquid and solid, $\Delta G^{L-S}$, that defines the melting temperature. This quantity can be very accurately determined for a classical system (noted as $\mathcal{C}$), and then the corresponding value in the *ab initio* system ($\mathcal{A}$) can be obtained using TI. To realize this approach, one needs a reasonably accurate semi-empirical potential under which the solid phase of interest is at least metastable, and a liquid structure is close to the *ab initio* one. Then accurate latent heat and the melting temperature can be obtained from classical MD simulation with the chosen semi-empirical potential and sufficiently large simulation cell and time. These results allow us to get the free energy difference between the solid and liquid phases, $\Delta G_\mathcal{C}^{L-S}$, associated with melting for the classical system from the Gibbs-Helmholtz equation. Finally, one can perform the TI from this classical system to the *ab initio* system to transform the classical $\Delta G_\mathcal{C}^{L-S}$ into the *ab initio* $\Delta G_\mathcal{A}^{L-S}$. The whole process does not involve any approximations, so the computed $\Delta G_\mathcal{A}^{L-S}$ should be exact. Moreover, since, by design, the classical potential already provides a close description of the solid and liquid phases compared to the *ab initio* models, minimal structural changes are expected during the TI. Harmonic and anharmonic contributions are inherently included in the SLC simulations, so this should provide very accurate results. Similar idea was first proposed by (Alfè et al., 2002a) to correct the melting properties obtained by classical coexistence simulations. In this study, we derived the formula for this approach and use it to compute the melting temperatures for the bcc and hcp iron under Earth's inner core pressures.

## 2 Methods

2.1 Formulas

Assuming the melting temperature $T_\mathcal{C}^m$ is known for the classical system, one can compute $\Delta G_\mathcal{C}^{L-S}$ at $T$ using the Gibbs-Helmholtz equations

$$\Delta G_\mathcal{C}^{L-S}(T) = -T \int_{T_\mathcal{C}^m}^{T} \frac{\Delta H_\mathcal{C}}{T^2} dT, \qquad (1)$$

where $\Delta H_\mathcal{C}$ is the latent heat of system $\mathcal{C}$, which can be directly determined from classical NPT MD simulations. The transformation from $\Delta G_\mathcal{C}^{L-S}(T)$ to $\Delta G_\mathcal{A}^{L-S}(T)$ can be obtained (see Supporting Information Text S1) as

$$\Delta G_\mathcal{A}^{L-S}(T) = \Delta G_\mathcal{C}^{L-S}(T) + f_{PV}(T) + f_{TI}(T). \qquad (2)$$

where $f_{PV}(T)$ is the contribution from the equation of state (EoS) difference between $\mathcal{A}$ and $\mathcal{C}$ systems, defined as

$$f_{PV}(T) = \left[P(V_\mathcal{A}^L - V_\mathcal{A}^S) - P(V_\mathcal{C}^L - V_\mathcal{C}^S)\right] - \left(\int_{V_\mathcal{C}^L}^{V_\mathcal{A}^L} P_\mathcal{C}^L(V)dV - \int_{V_\mathcal{C}^S}^{V_\mathcal{A}^S} P_\mathcal{C}^S(V)dV\right), \qquad (3)$$

where $V_\mathcal{A}^L$ (or $V_\mathcal{A}^S$) and $V_\mathcal{C}^L$ (or $V_\mathcal{C}^S$) are equilibrium volumes of liquid (or solid) at $P$ for system $\mathcal{A}$ and $\mathcal{C}$, respectively, $P_\mathcal{C}^L(V)$ and $P_\mathcal{C}^S(V)$ are equation of states of liquid and solid for system $\mathcal{C}$, respectively. $f_{TI}(T)$ term accounts for the TI difference between liquid and solid, which is defined as

$$f_{TI}(T) = \int_0^1 \langle U_{\mathcal{A}}^L - U_{\mathcal{C}}^L \rangle_{\lambda, NVT} d\lambda - \int_0^1 \langle U_{\mathcal{A}}^S - U_{\mathcal{C}}^S \rangle_{\lambda, NVT} d\lambda, \tag{4}$$

where $U_{\mathcal{A}}^L$ (or $U_{\mathcal{A}}^S$) and $U_{\mathcal{C}}^L$ (or $U_{\mathcal{C}}^S$) are the internal energy of liquid (or solid) for systems $\mathcal{A}$ and $\mathcal{C}$, respectively. $\langle \cdot \rangle_{\lambda, NVT}$ is the ensemble average of internal energy over configurations sampled in the canonical ensemble with the force field $U = (1-\lambda)U_{\mathcal{A}} + \lambda U_{\mathcal{C}}$. The subscript *NVT* means the constant conditions of $(V_{\mathcal{A}}^L, T)$ and $(V_{\mathcal{A}}^S, T)$ in liquid and solid simulations, respectively.

2.2 Simulation details

The classical MD simulations were performed using the Large-scale Atomic/Molecular Massively Parallel Simulator (LAMMPS) code (Thompson et al., 2022). The interatomic interaction was modeled using the semi-empirical potentials based on the embedded atom method (EAM)(Daw and Baskes, 1984). The system $\mathcal{C}$ was simulated using the semi-empirical potential developed in (Sun et al., 2022) and the system $\mathcal{C}'$ was simulated using the semi-empirical potential developed in the present study. In the NVT (constant number of atoms, volume, and temperature) simulations, the Nosé-Hoover thermostat (Nosé, 1984) was applied. In the NPT simulations, the Nosé-Hoover thermostat and barostat were applied. The damping time in the Nose-Hoover thermostat was set to $\tau = 0.01\ ps$. The time step of the simulation was $1.0\ fs$.

*Ab initio* calculations were performed with the Vienna *ab initio* simulation package (VASP) (Kresse and Furthmüller, 1996). The projected augmented-wave (PAW) method was used to describe the electron-ion interaction, and the generalized gradient approximation (GGA) in the Perdew-Burke-Ernzerhof (PBE) form was employed for the exchange-correlation energy functional. The electronic entropy at the high temperature was described by the Mermin functional (Mermin, 1965; Wentzcovitch et al., 1992). The electronic temperature in the Mermin functional is kept the same as the ionic temperature. In TI-MD from classical to *ab initio* systems, the force acting on each atom was $f = (1-\lambda)f_{\mathcal{A}} + \lambda f_{\mathcal{C}}$, where $f_{\mathcal{A}}$ and $f_{\mathcal{C}}$ were the forces generated by *ab initio* calculation and EAM potential, respectively. The *ab initio* forces computed from the VASP code were passed to the LAMMPS code on-the-fly with MD simulation. The Nosé-Hoover thermostat was employed to control the temperature, and a time step of $2.0\ fs$ was used to integrate Newton's equations of motion in the TI-MD.

Two PAW-PBE potentials, namely PAW8 and PAW16, were employed to balance the efficiency and accuracy of *ab initio* free energy calculations. PAW8 potential with 8 valence electrons ($3d^7 4s^1$) was used for AIMD and TI-MD simulations. PAW16 potential with 16 valence electrons ($3s^2 3p^6 3d^7 4s^1$) was used to improve the DFT accuracy with free energy perturbation (FEP). The plane-wave cutoff was 400 eV for PAW8 and 750 eV for PAW16. Supercells with 240, 250, and 250 atoms were used to simulate hcp, bcc, and liquid, respectively. The Γ point was used with PAW8 to sample the Brillouin zone in AIMD and TI-MD. A dense Monkhorst-Pack (Monkhorst and Pack, 1976) **k**-point mesh of 2 × 2 × 2 was adopted for all hcp, bcc, and liquid phases to achieve a high DFT accuracy in the FEP calculations.

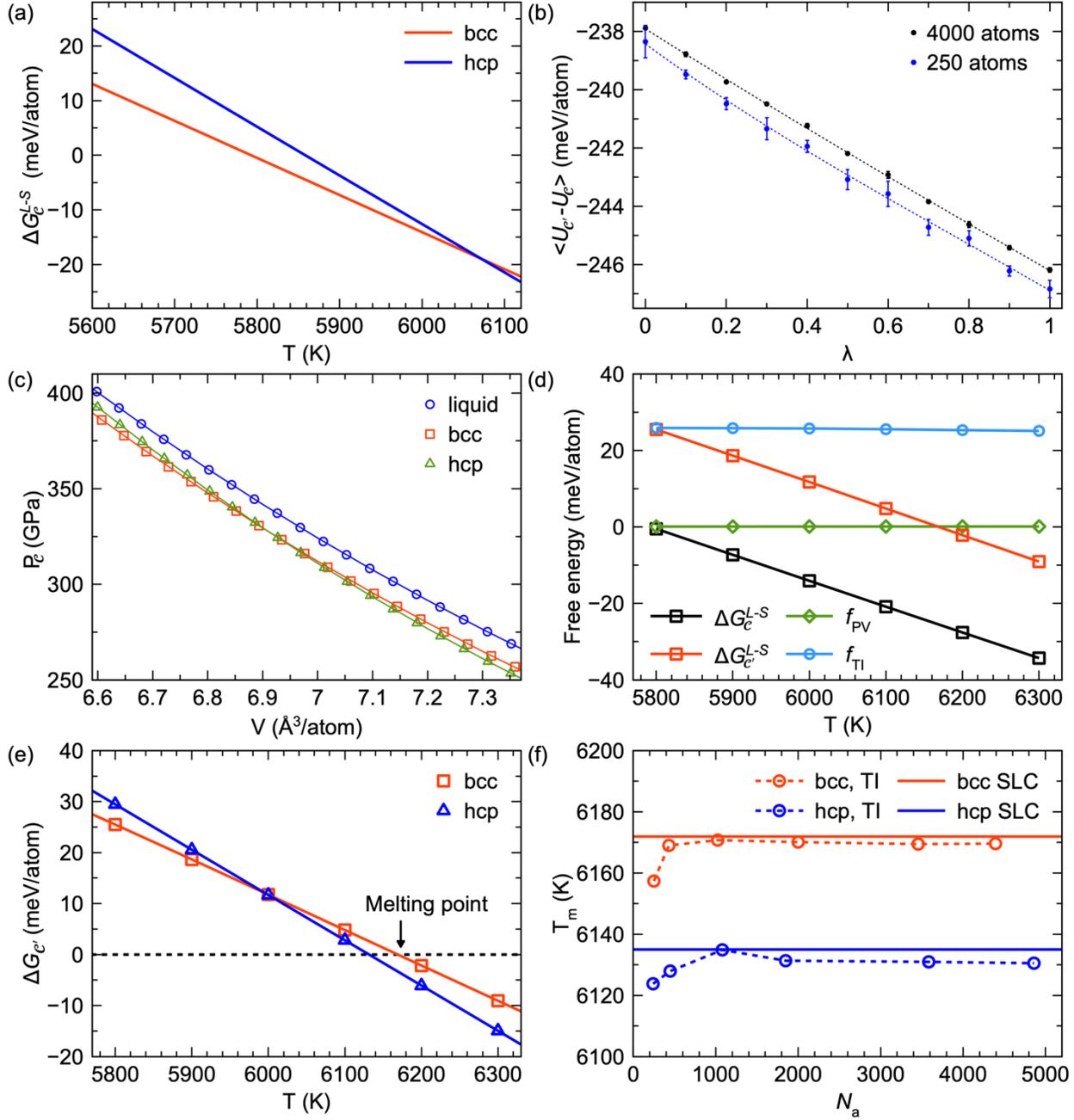

**Figure 1.** Melting temperature from $\mathcal{C} \to \mathcal{C}'$ calculations and size effect. (a) Gibbs free energy difference between the liquid and solid for bcc and hcp as a function of temperature in system $\mathcal{C}$ at 323GPa. (b) The integrand of TI along the thermodynamic path from $\mathcal{C}$ to $\mathcal{C}'$. The error bars are computed by repeating the simulation three times. The dotted line is a polynomial interpolation. (c) Equation of state in system $\mathcal{C}$ for liquid, bcc, and hcp at T=6000 K. The symbols are raw data obtained from MD simulations with 4000 atoms. The lines are the fitting of 3rd order Birch-Murnaghan EoS. (d) $\Delta G_{\mathcal{C}'}^{L-S}$ computed from $\Delta G_{\mathcal{C}}^{L-S}$ with the contribution of PV difference term and TI term, for bcc at 323 GPa using 4000 atoms. (e) The Gibbs free energy difference $\Delta G_{\mathcal{C}'}$ for bcc and hcp at 323 GPa. The solid line is an interpolation with 3rd polynomial fitting. (f) Size effect on the melting temperature measurement with TI simulations. The circles are melting temperatures measured from the TI with different numbers of atoms ($N_a$) in the simulations.

## 3 Results and Discussion

3.1 $\mathcal{C} \to \mathcal{C}'$ validation

Eqn. (2) is formally exact for sufficiently large simulation cells. However, such a condition is never satisfied for AIMD simulations. Since the size and time restrictions of MD simulations are not specific to the type of interatomic interaction (*ab initio* or classical), we can estimate the order of uncertainty by applying the approach to compute the melting temperature for another classical system $\mathcal{C}'$ where one can test from small to large simulation sizes. In this $\mathcal{C} \to \mathcal{C}'$ test, the starting system $\mathcal{C}$ is described by the Fe semi-empirical potential developed with EAM in (Sun et al., 2022), where the bcc and hcp Fe melting temperatures have been determined via large-scale SLC simulations as $T_{\mathcal{C}}^m$(bcc)=5,793 K and $T_{\mathcal{C}}^m$(hcp)= 5,858 K at P=323GPa. To build system $\mathcal{C}'$, we developed another Fe EAM potential in the same way as (Sun et al., 2022) except that we intentionally changed target melting temperatures. In $\mathcal{C}'$ system, the bcc and hcp melting temperatures measured from SLC simulations are $T_{\mathcal{C}'}^m$(bcc)=6,172 K and $T_{\mathcal{C}'}^m$(hcp)=6,135 K at 323 GPa. Note that the bcc phase has a higher $T_m$ than the hcp phase in the system $\mathcal{C}'$ while it is the opposite in the system $\mathcal{C}$. The melting temperature difference between bcc and hcp in the system $\mathcal{C}'$ is within 50 K, which is much smaller than the uncertainty of usual *ab initio* TI with small sizes using harmonic crystal and simple liquid as references. Therefore, this is a challenging testbed to examine the accuracy of the TI scheme.

To obtain $\Delta G_{\mathcal{C}'}^{L-S}$ we first computed the $\Delta G_{\mathcal{C}}^{L-S}(T)$ using the Gibbs-Helmholtz equation shown in Eqn. (1). The latent heat results are obtained from *NPT*-MD simulation with ~2,000 atoms. The results at 323 GPa are shown in Fig. 1(a). To compute the $f_{TI}(T)$ term in Eqn. (4), we performed TI-MD simulations for the system described by the hybrid Hamiltonian, $H = (1 - \lambda)H_{\mathcal{C}'} + \lambda H_{\mathcal{C}}$, where $H_{\mathcal{C}'}$ and $H_{\mathcal{C}}$ are the Hamiltonians of $\mathcal{C}'$ and $\mathcal{C}$ systems, respectively. The *NVT* ensemble was applied in the TI-MD simulation. The averaged internal energy differences between $\mathcal{C}'$ and $\mathcal{C}$ systems are shown as a function of $\lambda$ in Fig. 1(b), which defines the thermodynamic path from $\mathcal{C}'$ to $\mathcal{C}$. The path is smooth and almost linear, mainly because of the similarity between the two systems. The integrals are computed along the thermodynamic path via Gaussian quadrature. In order to evaluate the size effect, we compare the thermodynamic path using the simulation cells containing 250 and 4,000 atoms. It shows a systematic deviation between the two simulations. This size effect leads to an error of 0.7 meV/atom in the $f_{TI}(T)$. We will show later how this energy difference affects the calculated $T_m$. In addition, the error bar of the data from the 250-atom simulation is naturally much larger than the one obtained from the 4,000-atom simulation.

To compute $f_{PV}$, the first term in Eqn. (3) was directly obtained by measuring the equilibrated volumes of liquid and solids at 323 GPa for systems $\mathcal{C}'$ and $\mathcal{C}$. The second integral term in Eqn. (3) was obtained by measuring the equation of states systems $\mathcal{C}$. In Fig. 1(c), classical *NVT*-MD simulations are performed with large simulation cells (4032 atoms for hcp, 4394 atoms for bcc, 4000 atoms for liquid) and a serial of cell volumes to obtain $P_{\mathcal{C}}(V)$ near the pressure of the interest. These results were fitted to the 3rd order Birch-Murnaghan EoS so that the integral term in Eqn. (3) can be obtained.

Putting the $\Delta G_{\mathcal{C}}^{L-S}(T)$, $f_{TI}(T)$, and $f_{PV}(T)$ together via Eqn. (2), we computed the temperature-dependent Gibbs free energy $\Delta G_{\mathcal{C}'}^{L-S}(T)$ for system $\mathcal{C}'$. The contributions from $f_{TI}(T)$ and $f_{PV}(T)$ are shown in Fig. 1(d). The $f_{PV}(T)$ term is almost zero. The main factor in determining the $\Delta G_{\mathcal{C}'}^{L-S}(T)$ is the $f_{TI}(T)$ term. This term is almost independent of the temperature. The $\Delta G_{\mathcal{C}'}(T)$ for the bcc and hcp phases are shown in Fig. 1(e). We obtained $T_{\mathcal{C}'}^m(\text{hcp}, TI) = 6131.6\ K$ and $T_{\mathcal{C}'}^m(bcc, TI) = 6169.1\ K$. Therefore, we could reproduce the fact that the bcc is the most stable crystal phase in the system $\mathcal{C}'$ using the TI scheme. Moreover, the obtained values are very close to the melting temperatures measured by the SLC simulations, with differences of only 3.4 $K$ for the hcp phase and 2.8 $K$ for the bcc phase.

The size effect on the melting temperatures was examined in the $\mathcal{C} \rightarrow \mathcal{C}'$ case by performing the TI simulations using different numbers of atoms ($N_a$). Because the limit of the simulation size in the AIMD simulations only affects the measurement of $f_{TI}(T)$ term, the $\Delta G_{\mathcal{C}}^{L-S}(T)$ and $f_{PV}(T)$ terms are kept the same in this test. The $T_m$ values calculated from different simulations size are shown in Fig. 1(f), together with the $T_m$ data from SLC simulation for comparison. When the simulation size is larger than 1,000 atoms, the melting temperatures are almost independent of the simulation size and only show a deviation from the SLC results within 5 K. When the simulation size is reduced to ~ 250 atoms, which is the typical cell size of AIMD simulations, the uncertainty in $T_m$ increases to ~ 15 K. This is the consequence of 0.7 meV/atom difference in the $f_{TI}(T)$ between 250-atom and 4000-atom simulations, shown in Fig. 1(b). Compared to the TI calculation using harmonic crystal as the reference, the inaccuracy caused by the size effect is significantly reduced by the current TI scheme. This can be rationalized as follows. The size effect is associated with the cutting off longwave phonons in a small simulation cell of crystal phases (the size effect is much less important for liquid models). When a TI is performed from a system described by a semi-empirical potential to the *ab initio* system, the effect of the simulation cell on the TI results is only associated with the difference in the phonon distributions between the classical and *ab initio* systems, which is supposed to be small if the semi-empirical potential was well fitted to *ab initio* results. The major contribution of longwave phonons to the free energy difference between the solid and liquid phases is incorporated in reference systems via large-scale SLC simulations. On the contrary, when the TI is performed from a harmonic crystal to the *ab initio* system with a small simulation size, the effect of longwave phonons is ignored because the reference harmonic crystal does not contain the longwave phonon contribution.

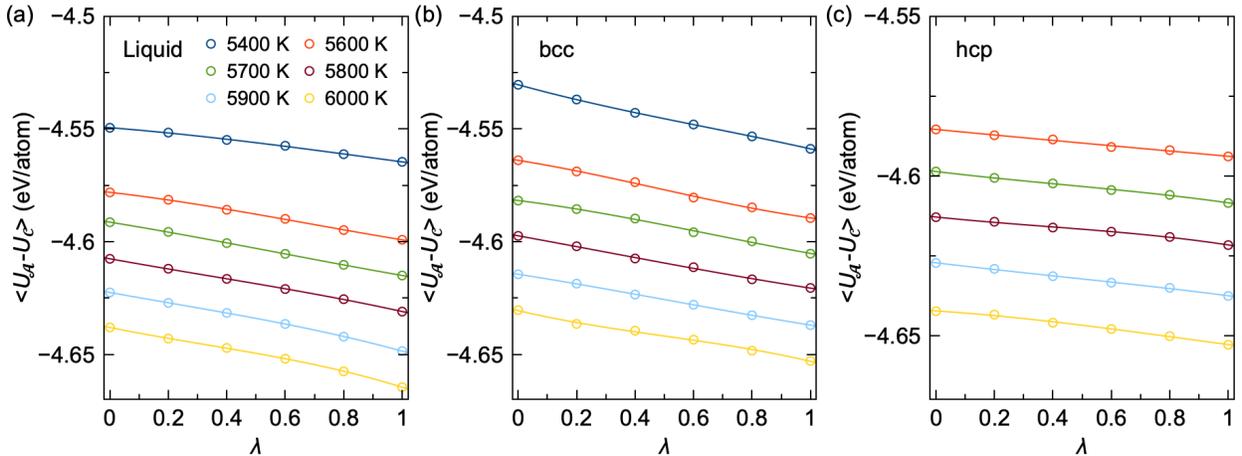

**Figure 2.** $\mathcal{C} \to \mathcal{A}$ TI calculations at P=323 GPa. (a)-(c) show the averaged energy differences between *ab initio* system $\mathcal{A}$ and classical system $\mathcal{C}$ in TI-MD simulations for liquid, bcc, and hcp, respectively. The solid lines are the polynomial interpolations.

3.2 Classical to *ab initio* transformation ($\mathcal{C} \to \mathcal{A}$)

We now compute *ab initio* melting temperatures for hcp and bcc. It requires $\Delta G_{\mathcal{C}}^{L-S}(T)$, $f_{TI}(T)$, and $f_{PV}(T)$ to obtain the $\Delta G_{\mathcal{A}}^{L-S}(T)$ via Eqn. (2). $\Delta G_{\mathcal{C}}^{L-S}(T)$ is the same as the one computed in the $C \to C'$ case. The main task is to compute the $f_{TI}(T)$ from classical system $C$ to *ab initio* system $\mathcal{A}$.

The *ab initio* TI-MD was performed for liquid, hcp, and bcc in the temperature range from 5,400K to 6,000K. The energy differences between *ab initio* system $\mathcal{A}$ and classical system $C$, i.e., $\Delta U = \langle U_{\mathcal{A}} - U_C \rangle$, are averaged from TI-MD over 20 *ps* and shown for three phases in Fig. 2. Just like in the $C \to C'$ case, all the $\Delta U$ show smooth transitions from the classical system ($\lambda = 0$) to the *ab initio* system ($\lambda = 1$). These thermodynamic paths are almost linear, similar to the $C \to C'$ results in Fig. 1(b). The $f_{PV}(T)$ terms are computed with the *ab initio* equilibrium volumes of hcp, bcc, and liquid and the classical equation of states $P_C(V)$. The $f_{PV}(T)$ terms is also almost zero, similar to the one in $C \to C'$ calculations. With these quantities, $\Delta G_{\mathcal{A}}^{L-S}(T)$ was obtained and shown in Fig. 3. The *ab initio* melting temperatures were obtained from the conditions $\Delta G_{\mathcal{A}}^{L-S}(T_m) = 0$. The $T_m$ results are $5848 \pm 15$ K for hcp and $5632 \pm 15$ for bcc at 323GPa, while $6094 \pm 15$ K for hcp and $5850 \pm 15$ K for bcc at 360 GPa. The uncertainty of 15 K was estimated based on the $C \to C'$ test of the finite size effect shown in Fig. 1(f). These results suggest the hcp always has a high melting temperature than bcc from 323 GPa to 360 GPa. The free energy differences between the bcc and hcp phases are quite small, ~13 meV/atom at the hcp melting point. This is almost the uncertainty of the conventional TI method (Alfè et al., 2001). Therefore, the present TI with classical potential is a necessity to distinguish the free energy difference between hcp and bcc iron under core conditions.

3.3 DFT accuracy

While we have addressed the size effect with the TI from classical to *ab initio* system, there still exists uncertainty within *ab initio* calculations. So far, the *ab initio* calculations in the TI were performed at the PAW8 potential level (8 valence electrons). However, Fe's semi-core electron orbitals can contribute to the metallic bonding and affect the melting properties under the Earth's core pressures (Alfè et al., 2002b). This effect can be evaluated by PAW potentials with more semi-core electrons treated as valence electrons, such as PAW16 potential (16 valence electrons) (Sun et al., 2018). However, due to the intrinsic cubic scaling of electron number in the DFT, PAW16 increases the computational cost by a factor of ~10, such that it is difficult to implement PAW16 in the present TI scheme. Recently, the FEP method was demonstrated to be an efficient approach for transforming the free energy data from Fe's PAW8 to PAW16 (Sun et al., 2018). Therefore, we employed the FEP method to improve the DFT accuracy (see Supporting

Information) at three temperatures for both 323 GPa and 360 GPa. The free energy changes associated with replacing PAW8 with PAW16 are shown for both hcp and bcc in Supporting Information Fig. S1. It leads to the stabilization of the hcp and bcc phases with respect to the liquid phase by 30-50 meV/atom. Therefore, the melting temperatures of the hcp and bcc phases are increased compared to the PAW8 results, as shown in Fig. 3. The error of the FEP method was estimated as 3 meV/atom (Sun et al., 2018). This error propagates to the melting temperatures and results in a larger uncertainty compared to the original TI results. The melting temperatures obtained with the PAW16 potential are $6357 \pm 45$ K for hcp and $6168 \pm 80$ K for bcc at 323GPa, while $6692 \pm 45$ K for hcp and $6519 \pm 80$ K for bcc at 360 GPa. Within the uncertainty, the hcp phase still shows a higher melting temperature than the bcc phases. The free energy differences between the bcc and hcp phases are ~8 meV/atom at the hcp melting point. Therefore, the relative stability between hcp and bcc is qualitatively similar for the PAW8 and PAW16 potentials. We note that the effect of semi-core states on PAW8 can also be corrected by pair potential, as demonstrated by (Alfè et al., 2002b).

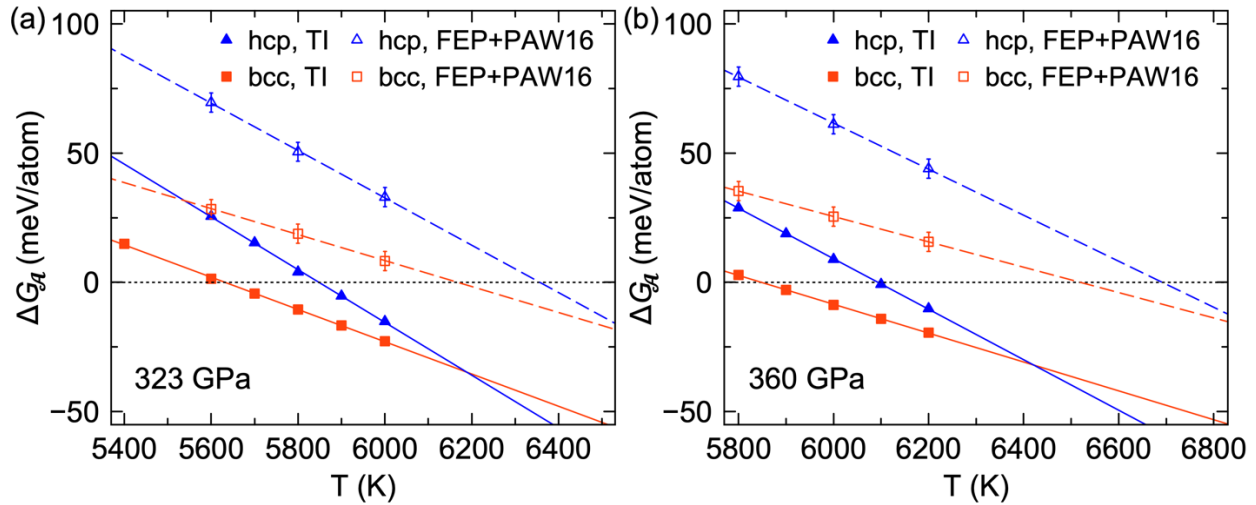

**Figure 3.** The *ab initio* Gibbs free energy difference $\Delta G_{\mathcal{A}}^{L-S}$ for bcc and hcp. (a) 323GPa and (b) 360GPa. The lines are linear interpolations of the data points. The intersection with the dotted line ($\Delta G_{\mathcal{A}}^{L-S} = 0$) defines the melting temperature.

3.4 Geophysical impact

The obtained *ab initio* melting temperatures are compared with the data from the literature in Fig. 4. While it shows rich melting temperature data for the hcp phase, the bcc data are very limited. The only *ab initio* datum of bcc melting point is from (Bouchet et al., 2013). The calculation was performed by 256-atom SLC simulations, which led to large uncertainties in both pressure and temperature. Nevertheless, the results are consistent with the present PAW16 results for bcc. For *ab initio* hcp data, the only PAW8 result is from (Sun et al., 2018), which is consistent with our PAW8 data within the uncertainty. Our PAW16 data of hcp is also consistent with 980-atom SLC simulations (Alfè, 2009). The results are close to Alfè *et al.*'s TI calculations (Alfè et al., 2002b) and Monte Carlo results (Sola and Alfè, 2009). The data also agrees with recent experimental results within their uncertainty (one of them (Li et al., 2020) shown in Fig. 4). Overall, the present calculations are consistent with previous results for the hcp phase with the same PAW potentials, which validates our TI methods. Moreover, it provides a clear distinction between the hcp and bcc phases under the same DFT accuracy.

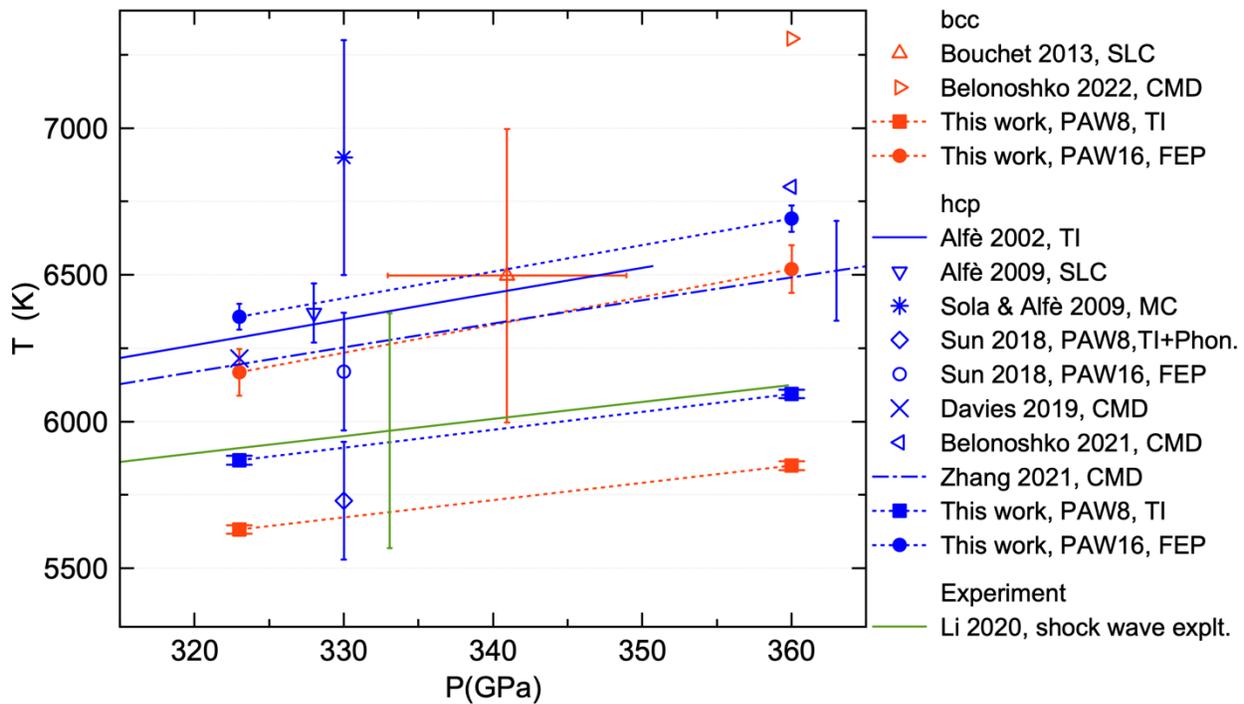

**Figure 4.** Melting temperatures of bcc and hcp phases. The literature data includes Alfè 2002 by TI (Alfè et al., 2002b), Alfè 2009 by *ab initio* SLC simulations (Alfè, 2009), Sola & Alfè 2009 by Monte Carlo (Sola and Alfè, 2009), Bouchet 2013 by *ab initio* SLC simulations (Bouchet et al., 2013), Sun 2018 by TI, phonon quasiparticle and FEP (Sun et al., 2018), Davies 2019 by classical MD (Davies et al., 2019), Belonoshko 2021 & Belonoshko 2022 by classical MD (Belonoshko et al., 2022a, 2021) and Zhang 2021 by classical MD with a machine-learning potential (Zhang et al., 2020). The experimental curve is from Li 2020 (Li et al., 2020) with the extrapolation of shock wave data.

Another group of data is from the classical MD with EAM (Belonoshko et al., 2022a, 2021; Davies et al., 2019) and machine-learning potentials (Zhang et al., 2020). While these results depend on the employed potentials, they mostly show decent consistency with present PAW16 data on the hcp melting temperatures. The largest deviation is on the bcc phase from (Belonoshko et al., 2022a, 2021). It leads to bcc's melting temperature of ~780 K higher than our PAW16 value at 360 GPa; more importantly, it indicates that the hcp melting temperature is much lower than that for the bcc phase. Such deviation should be due to the employed semi-empirical potential, which was fitted to 0 K DFT results, and the calculated free energy contained an *ad hoc* correction of the electronic entropy effect (Belonoshko et al., 2022a, 2021).

According to the present TI and FEP calculations, the hcp phase has higher melting temperatures than the bcc phase under both pressures of the inner core boundary and the inner core center. Therefore, the hcp phase should be the stable phase for pure Fe at inner-core conditions. While there are increasing evidence of the bcc phase in the inner core (Belonoshko et al., 2022b; Ghosh and Zhang, 2022), our *ab initio* calculations, regardless of the PAW potential employed, suggest that the stability of the bcc phase should not be due to Fe itself. The free energy difference between bcc and hcp is tiny, ~10 meV/atom at 323 GPa. Such a small free energy difference could be altered by alloying with other elements, especially if the alloy's configuration entropy becomes significant. For instance, a recent experimental work reported the hcp-B2 coexistence in Fe-Ni-Si alloy under core conditions (Ikuta et al., 2021). Therefore, the effect of nickel and light elements must be taken into account to get a complete picture of the crystalline structure of the solid inner core phase.

## 4. Conclusion

We developed a TI scheme to compute the free energy difference between solid and liquid. This scheme uses the empirical potential as the reference system, providing a smooth TI calculations transition. The $T_m$ of the target system can be computed without calculating the absolute free energy. The test between two classical systems ($\mathcal{C} \to \mathcal{C}'$) shows that the obtained $T_m$ is consistent with that from large-scale SLC simulations, which provides a good validation for the method. The size effect in the current TI scheme is small, on the order of ~15 K difference between a 250-atom cell and a 4,000-atom cell. Based on this scheme, the *ab initio* melting temperatures of hcp and bcc iron are calculated at the inner core boundary and the core center pressures. The hcp always displays a higher $T_m$ than the bcc. Thus, hcp should be the stable phase of pure iron at the inner core conditions. However, the free energy difference between the two phases is only of the order of 10s meV/atom. Future work should include the effect of nickel and light elements on the free energy, which is necessary to simulate the inner core structure.

**Acknowledgments**

Work at Iowa State University and Columbia University was supported by the National Science Foundation awards EAR-1918134 and EAR-1918126. We acknowledge the computer resources from the Extreme Science and Engineering Discovery Environment (XSEDE), which is supported by the National Science Foundation grant number ACI-1548562. B. D. is supported by JSPS KAKENHI Grant Number JP21K14656. Molecular dynamics simulations were supported by the Numerical Materials Simulator supercomputer at the National Institute for Materials Science (NIMS).

**Open Research**

Data Availability Statement: The authors comply with the AGU's data policy, and the datasets of this paper are available at zenodo: https://doi.org/10.5281/zenodo.7407428

Supporting Information for

# *Ab initio* melting temperatures of bcc and hcp iron under the Earth's inner core condition

Yang Sun[1,2,3], Mikhail I. Mendelev[4], Feng Zhang[3], Xun Liu[5], Bo Da[5], Cai-Zhuang Wang[3]

Renata M. Wentzcovitch[2,6,7] and Kai-Ming Ho[3]

[1]Department of Physics, Xiamen University, Xiamen, Fujian 361005, China
[2]Department of Applied Physics and Applied Mathematics, Columbia University, New York, NY 10027, USA
[3]Department of Physics, Iowa State University, Ames, IA 50011, USA
[4]Intelligent Systems Division, NASA Ames Research Center, Moffett Field, CA 94035, USA
[5]Research and Services Division of Materials Data and Integrated System, National Institute for Materials Science, Ibaraki 305-0044, Japan.
[6]Department of Earth and Environmental Sciences, Columbia University, New York, NY 10027, USA
[7]Lamont–Doherty Earth Observatory, Columbia University, Palisades, NY 10964, USA

**Contents of this file:**

Text S1-2
Figure S1

**Introduction**

This Supporting Information provides the derivation of thermodynamic intergration formula in Text S1, details of free energy perturbation in Text S2, and the data of free energy perturbation in Fig. S1.

**Text S1. Formulas for $\mathcal{C} \to \mathcal{A}$ thermodynamic intergration**

In this section, we provide the derivation of formulas transforming classical T$_m$ and solid-liquid free energy difference to *ab initio* ones. Let's consider two systems $\mathcal{A}$ and $\mathcal{C}$, described by *ab initio* and classical force fields, respectively. We assume the melting temperature $T_\mathcal{C}^m$ are known for the classical system (calculated by large solid-liquid coexistence simulations in practice). In the following, we derive the equation to obtain the free energy difference and $T_m$ for system $\mathcal{A}$, i.e. $\Delta G_\mathcal{A}^{L-S}$ and $T_\mathcal{A}^m$. We start with the definition:

$$\Delta G_\mathcal{A}^{L-S}(T) = F_\mathcal{A}^L(V_\mathcal{A}^L, T) - F_\mathcal{A}^S(V_\mathcal{A}^S, T) + P(V_\mathcal{A}^L - V_\mathcal{A}^S), \tag{S1}$$

where $P$ and $T$ are pressure and temperature, $V_\mathcal{A}^L$ and $V_\mathcal{A}^S$ are equilibrium volumes and $F_\mathcal{A}^L(T)$ and $F_\mathcal{A}^S(T)$ are Helmholtz free energies of liquid and solid in system $\mathcal{A}$ at $(P, T)$, respectively. Performing the thermodynamic intergration (TI) from the classical system $\mathcal{C}$ for liquid and solid, one can compute the Helmholtz free energies as

$$\begin{aligned}F_\mathcal{A}^L(V_\mathcal{A}^L, T) - F_\mathcal{C}^L(V_\mathcal{A}^L, T) &= \int_0^1 \langle U_\mathcal{A}^L - U_\mathcal{C}^L \rangle_{\lambda, NVT} d\lambda \\ F_\mathcal{A}^S(V_\mathcal{A}^S, T) - F_\mathcal{C}^S(V_\mathcal{A}^S, T) &= \int_0^1 \langle U_\mathcal{A}^S - U_\mathcal{C}^S \rangle_{\lambda, NVT} d\lambda\end{aligned}, \tag{S2}$$

where $\langle \cdot \rangle_{\lambda, NVT}$ is the ensemble average of internal energy over configurations sampled in the canonical ensemble with the force field $U = (1 - \lambda)U_\mathcal{A} + \lambda U_\mathcal{C}$. The subscript *NVT* means the constant conditions of $(V_\mathcal{A}^L, T)$ and $(V_\mathcal{A}^S, T)$ in liquid and solid simulations, respectively. We note the *ab initio* internal energy, $U_\mathcal{A}$, includes the electronic entropy contribution. For simplicity's sake, we denote the integral terms in Eqn. (2) as $F_{TI}^L(V_\mathcal{A}^L, T) = \int_0^1 \langle U_\mathcal{A}^L - U_\mathcal{C}^L \rangle_{\lambda, NVT} d\lambda$ and $F_{TI}^S(V_\mathcal{A}^S, T) = \int_0^1 \langle U_\mathcal{A}^S - U_\mathcal{C}^S \rangle_{\lambda, NVT} d\lambda$. Then Eqn. (2) can be rewritten as

$$F_\mathcal{A}^L(V_\mathcal{A}^L, T) - F_\mathcal{A}^S(V_\mathcal{A}^S, T) = F_{TI}^L(V_\mathcal{A}^L, T) - F_{TI}^S(V_\mathcal{A}^S, T) + F_\mathcal{C}^L(V_\mathcal{A}^L, T) - F_\mathcal{C}^S(V_\mathcal{A}^S, T). \tag{S3}$$

The last two terms in Eqn. (3) require the free energy of the classical system $\mathcal{C}$. Given the calculated melting temperature $T_\mathcal{C}^m$, one can accurately compute the Gibbs free energy difference between liquid and solid $\Delta G_\mathcal{C}^{L-S}$ at $T$ using the Gibbs-Helmholtz equations

$$\Delta G_{\mathcal{C}}^{L-S}(T) = -T \int_{T_{\mathcal{C}}^{m}}^{T} \frac{\Delta H_{\mathcal{C}}}{T^2} dT, \tag{S4}$$

where $\Delta H_{\mathcal{C}}$ is the latent heat of system $\mathcal{C}$, which can be directly determined from classical *NPT* MD simulations. As $\Delta G_{\mathcal{C}}^{L-S}(T) = G_{\mathcal{C}}^{L}(T) - G_{\mathcal{C}}^{S}(T)$, one can write it as

$$\Delta G_{\mathcal{C}}^{L-S}(T) = F_{\mathcal{C}}^{L}(V_{\mathcal{C}}^{L}, T) - F_{\mathcal{C}}^{S}(V_{\mathcal{C}}^{S}, T) + P(V_{\mathcal{C}}^{L} - V_{\mathcal{C}}^{S}), \tag{S5}$$

where $V_{\mathcal{C}}^{L}$ and $V_{\mathcal{C}}^{S}$ are the equilibrium volumes of classical liquid and solid under the condition of $(P, T)$. Note $V_{\mathcal{C}}^{L}$ and $V_{\mathcal{C}}^{S}$ are not necessarily the same as $V_{\mathcal{A}}^{L}$ and $V_{\mathcal{A}}^{S}$ owing to the difference between classical and *ab initio* force fields. Since $P = -\frac{\partial F}{\partial V}|_T$, we can write

$$\begin{aligned} F_{\mathcal{C}}^{L}(V_{\mathcal{A}}^{L}, T) - F_{\mathcal{C}}^{L}(V_{\mathcal{C}}^{L}, T) &= -\int_{V_{\mathcal{C}}^{L}}^{V_{\mathcal{A}}^{L}} P_{\mathcal{C}}^{L}(V) dV \\ F_{\mathcal{C}}^{S}(V_{\mathcal{A}}^{S}, T) - F_{\mathcal{C}}^{S}(V_{\mathcal{C}}^{S}, T) &= -\int_{V_{\mathcal{C}}^{S}}^{V_{\mathcal{A}}^{S}} P_{\mathcal{C}}^{S}(V) dV \end{aligned}, \tag{S6}$$

where $P_{\mathcal{C}}^{L}(V)$ and $P_{\mathcal{C}}^{S}(V)$ are the equation of states of liquid and solid in the classical system at $T$. Combining both Eqns. (6) we obtain

$$F_{\mathcal{C}}^{L}(V_{\mathcal{A}}^{L}, T) - F_{\mathcal{C}}^{S}(V_{\mathcal{A}}^{S}, T) = F_{\mathcal{C}}^{L}(V_{\mathcal{C}}^{L}, T) - F_{\mathcal{C}}^{S}(V_{\mathcal{C}}^{S}, T) - \left( \int_{V_{\mathcal{C}}^{L}}^{V_{\mathcal{A}}^{L}} P_{\mathcal{C}}^{L}(V) dV - \int_{V_{\mathcal{C}}^{S}}^{V_{\mathcal{A}}^{S}} P_{\mathcal{C}}^{S}(V) dV \right). \tag{S7}$$

Now combining Eqn. (3), (5), and (7), we can write Eqn. (1) as

$$\Delta G_{\mathcal{A}}^{L-S}(T) = \Delta G_{\mathcal{C}}^{L-S}(T) + F_{TI}^{L}(V_{\mathcal{A}}^{L}, T) - F_{TI}^{S}(V_{\mathcal{A}}^{S}, T) + [P_0(V_{\mathcal{A}}^{L} - V_{\mathcal{A}}^{S}) - P_0(V_{\mathcal{C}}^{L} - V_{\mathcal{C}}^{S})] - \left( \int_{V_{\mathcal{C}}^{L}}^{V_{\mathcal{A}}^{L}} P_{\mathcal{C}}^{L}(V) dV - \int_{V_{\mathcal{C}}^{S}}^{V_{\mathcal{A}}^{S}} P_{\mathcal{C}}^{S}(V) dV \right). \tag{S8}$$

We define the TI term $f_{TI}(T)$ and the PV difference term $f_{PV}(T)$ as

$$f_{TI}(T) = F_{TI}^{L}(V_{\mathcal{A}}^{L}, T) - F_{TI}^{S}(V_{\mathcal{A}}^{S}, T), \tag{S9}$$

$$f_{PV}(T) = [P(V_{\mathcal{A}}^{L} - V_{\mathcal{A}}^{S}) - P(V_{\mathcal{C}}^{L} - V_{\mathcal{C}}^{S})] - \left( \int_{V_{\mathcal{C}}^{L}}^{V_{\mathcal{A}}^{L}} P_{\mathcal{C}}^{L}(V) dV - \int_{V_{\mathcal{C}}^{S}}^{V_{\mathcal{A}}^{S}} P_{\mathcal{C}}^{S}(V) dV \right). \tag{S10}$$

Now Eqn. (8) can be simplified as

$$\Delta G_{\mathcal{A}}^{L-S}(T) = \Delta G_{\mathcal{C}}^{L-S}(T) + f_{TI}(T) + f_{PV}(T). \tag{S11}$$

Using $\Delta G_{\mathcal{A}}^{L-S}(T)$ one can compute the melting temperature under the condition of $\Delta G_{\mathcal{A}}^{L-S}(T_{\mathcal{A}}^{m}) = 0$. To this end, the $f_{TI}(T)$ term requires TI simulation from system $\mathcal{C}$ to system $\mathcal{A}$. The $f_{PV}(T)$ term requires equilibrium volume $V_{\mathcal{A}}^{L}$ and $V_{\mathcal{A}}^{S}$ at $P$ for system $\mathcal{A}$ and equation of state $P_{\mathcal{C}}^{L}(V)$ and $P_{\mathcal{C}}^{S}(V)$, which can all be directly computed from *ab initio* and classical MD simulations.

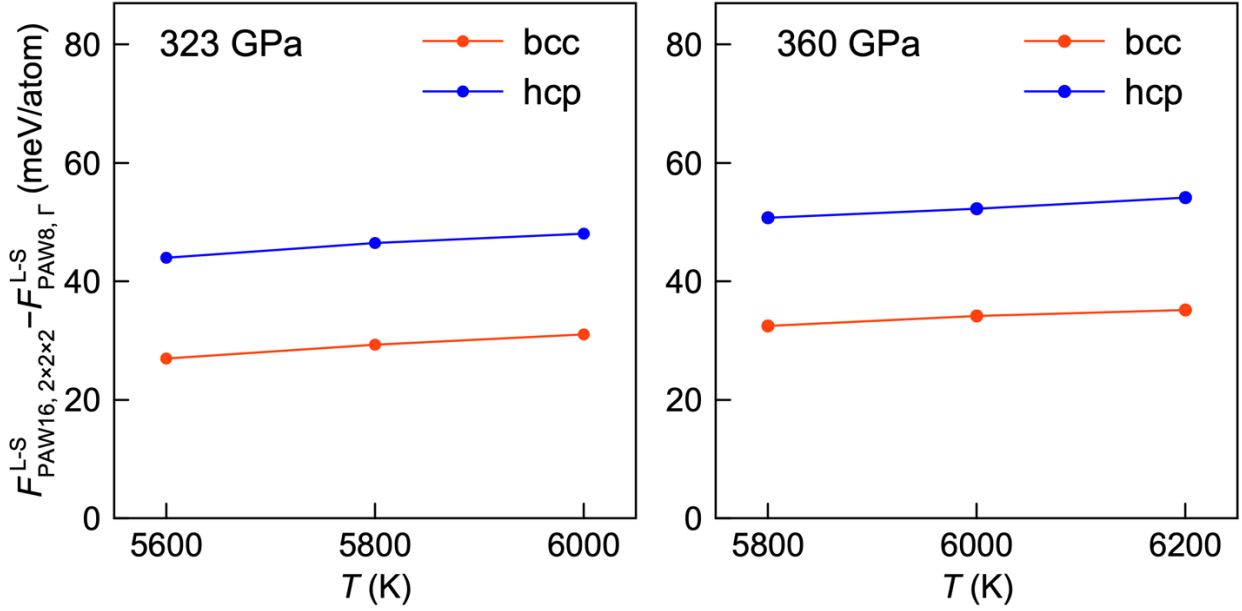

**Figure S1** Free energy difference between high-accuracy calculations (PAW16 potential and $2 \times 2 \times 2$ **k**-point mesh) and high-efficiency calculations (PAW8 potential and $\Gamma$ point).

**Text S2. Free energy perturbation**

The intensive *ab initio* calculations involved in the TI require a high-efficiency setup of density functional theory (DFT) calculations. The obtained free energy data from TI can further be improved to a higher DFT accuracy with the free energy perturbation (FEP) method (Zwanzig, 2004). In the FEP method, the free energy difference between system A and system B can be computed by the Zwanzig equation,

$$F_A - F_B = -k_B T \ln \left\langle \exp\left(-\frac{U_A - U_B}{k_B T}\right) \right\rangle_B, \tag{S12}$$

where $U_A$ and $U_B$ are the energy of the same atomic configuration computed with A and B interactions, respectively. $\langle \cdot \rangle_B$ is the ensemble average of system B. Equation (S12) is formally exact, while the convergence of the ensemble average depends on the similarity between A and B systems. In practice, one can use the trajectory from ab initio molecular dynamics of a high-efficiency setup and recompute the energy of random snapshots with the high-accuracy DFT setup. Equation (S12) is applicable if the fluctuation of $(U_A - U_B)$ is much smaller than $k_B T$. In our study, 50 MD snapshots were used to perform the ensemble average. The fluctuations of the energy difference between PAW8 and PAW16 calculations are in the range of 3-8 meV/atom for hcp, bcc, and liquid phases, which is much smaller than the $k_B T$ of ~ 500 meV/atom at ~ 6000 K. Therefore, FEP method is applicable for the present system.